\documentclass[aps,prl,showpacs,twocolumn]{revtex4-1}
\usepackage{graphicx, hyperref}
\usepackage{amsmath, amssymb}
\usepackage{textcomp}
\begin{document}

\title{Magnetic friction: From Stokes to Coulomb behavior}

\author{Martin P. Magiera}
\email{martin.magiera@uni-due.de} 
\author{Sebastian Angst}
\author{Alfred Hucht}
\author{Dietrich E. Wolf}
\affiliation{Faculty of Physics and CeNIDE, University of Duisburg-Essen,
  D-47048 Duisburg, Germany}

\pacs{75.10.Pq, 75.10.Hk, 68.35.Af}
\date{\today}

\begin{abstract}
  We demonstrate that in a ferromagnetic substrate, which is
  continuously driven out of equilibrium by a field moving with
  constant velocity $v$, at least two types of friction may occur when
  $v$ goes to zero: The substrate may feel a friction force
  proportional to $v$ (Stokes friction), if the field changes on a
  time scale which is longer than the intrinsic relaxation time.
  On the other hand, the friction force may become independent of $v$
  in the opposite case (Coulomb friction). These observations are analogous
  to e.g.\ solid friction. The effect is demonstrated in
  both, the Ising (one spin dimension) and the Heisenberg model (three
  spin dimensions), irrespective which kind of dynamics (Metropolis
  spin-flip dynamics or Landau-Lifshitz-Gilbert precessional dynamics)
  is used. For both models the limiting case of Coulomb friction can
  be treated analytically.  Furthermore we present an empiric
  expression reflecting the correct Stokes behavior and therefore
  yielding the correct cross-over velocity and dissipation.
\end{abstract}

\maketitle

Friction phenomena, despite their huge importance in everyday life,
are still not fully understood. Different friction mechanisms are
possible, leading to different dependencies of the friction forces on
the driving velocity. Microscopically, one often assumes Stokes-like
friction, i.e. a linear velocity dependence. However, this
atomistic view is in conflict with Coulomb friction at the interface
between solids, because it approaches a
nonzero absolute value in the limit of small velocities, independent
of the materials and their surface conditions \cite{Mate87, Liu94,
  Zwoerner98, Bennewitz00, Gnecco00, Mueser02}.  A possible solution
was offered by the simple model developed by Prandtl
and Tomlinson, in which a stick-slip
instability was responsible for Coulomb friction \cite{Prandtl1928,
  Tomlinson1929}. They suggested a surface atom to be coupled by a
spring of stiffness $k$ to a slider which moves with constant velocity
$v$. The atom interacts with the surface via a periodic potential and
experiences a viscous friction force proportional to its velocity
$\dot x$. If $k$ is sufficiently small with respect to the potential
height, the atom first gets stuck in the potential minima and slips
when the tension gets large enough. The slip motion 
$\dot x$ does not depend on the slider's velocity $v$, and one
observes Coulomb friction. However, when $k$ is large with respect to
the potential height, the atom moves with the slider's velocity and
the friction force is Stokesian. The cross-over from one regime to the
other has been studied recently \cite{Mueser11}.

What remains a puzzle, however, is that 
Coulomb friction is a far more general phenomenon than one might
expect from the Prandtl-Tomlinson model, which is formulated in terms
of elastic forces
in a periodic potential. For example, Stokes as well as Coulomb behavior
has also been observed for magnetic friction, where elastic forces
are absent. Being guided by a detailed investigation of the crossover
between both types of magnetic friction, a unifying principle can be
formulated that applies to the magnetic as well as to the elastic case.
 
So far, magnetic friction has been studied in two different types of models.
Ising models with single-spin-flip dynamics,
where two half spaces move with respect to each other,
yield Coulomb friction \cite{Kadau08, Hucht09, Hilhorst11}. Analogous
results have been obtained in the Potts model \cite{Igloi11}. On the
other hand, a magnetic dipole scanning a Heisenberg surface showed
Stokesian friction \cite{Fusco08, Magiera09, Magiera11,
  NIC-Proceeding} (always provided the velocity is not too large). 

Recently a work has been published, in which
a point-like magnetic perturbation moves through an Ising model \cite{Demery10, Demery10_2}. The authors claim to have observed Stokes
friction, which is in conflict with our results for similar models
\cite{Kadau08, Hucht09}. 
Here we present an
explanation of this discrepancy
and clarify, under
what conditions Stokes respectively Coulomb friction occurs.

The systems studied in Refs.~\cite{Kadau08, Hucht09, Hilhorst11, Igloi11} have in common
that the motion occurs in a discretized way: The system is at rest for
a certain number $a/v$ of Monte Carlo sweeps (MCS), after which one half
space is moved by one lattice constant $a$. Accordingly we have a
periodic excitation and relaxation procedure, where excitation
is fast (happens in between two subsequent spin flip attempts), 
whereas relaxation extends over $a/v$ MCS.  
By contrast, in \cite{Fusco08, Magiera09, Magiera11, NIC-Proceeding} excitation is slow, because due to the
dipole-dipole interaction a substrate spin feels the approaching tip a
long time in advance. 

\begin{figure}[bt]
\centering
\includegraphics[width=.9\columnwidth]{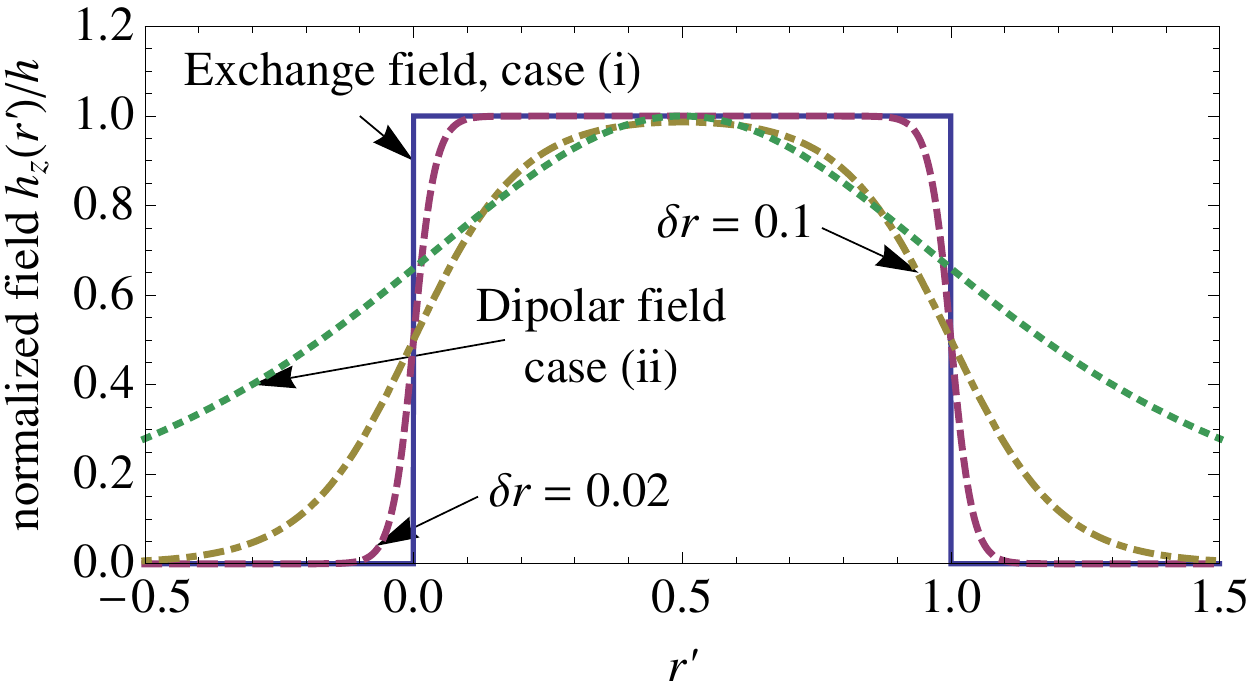}
\caption{\label{fig:skizze_felder}
The dynamics in the different studies can be mapped on a time
dependent field with amplitude $h_z(r')$ (here normalized by its
maximal value $h$) interacting with the
spins positioned at integer sites.
The discrete motion in the Ising/Potts model then corresponds to
a step-function, which may be treated as a fixed spin interacting via
exchange with one partner on the chain. The amplitude of a dipole
field is sketched for comparison. The
field used in this work may be tuned by adjusting the parameter
$\delta r$ from one limiting case to the other.
}
\end{figure}

Now we present a simple $1d$ model that interpolates between both cases: 
We consider a position dependent field $h_z(r')$, which is moved
continuously with constant velocity $v$. $r$ is given in units of $a$,
and $v=\dot r$. Then the discrete
motion can be modeled as a step function, as shown in
Fig.~\ref{fig:skizze_felder} as solid line. 
For a certain time $1/v$ exactly one spin is exposed
  to the field with constant amplitude until the field reaches the next
  spin. 
 Additionally the amplitude of the dipole field used in
\cite{Fusco08, Magiera09, Magiera11, NIC-Proceeding} is plotted. From
Ref.~\cite{Magiera09} we know that for this case the adjustment of the
spins with respect to the moved field happens in an adiabatic way, or in other
words the time scale of relaxation is below that of the excitation.
To generalize these setups, we  consider a field with steepness
$\delta r \ll 1$,
\begin{equation}
   h_{z}(r') =\frac{h}{\big (e^{-\frac{r'}{\delta r}}+1
    \big) \big (e^{-\frac{1-r'}{\delta r}}+1 \big )},
\label{eq:H}
\end{equation}
which may be tuned from the step-like field ($\delta r=0$, now called
limiting case (i)) to a slowly varying field ($\delta r \approx
0.1$, case (ii)).
By shifting this field according to $r'=r-v t$ \footnote{A prime denotes a quantity in the field's frame of
  reference, otherwise in the laboratory
  frame. I.e., at $1'$ ($0'$) is the front (rear) inflection point
  of $h$.} we can directly
influence the time scale at which the
excitation at a fixed position $r$ occurs, $\tau_\mathrm{switch} \propto \delta r/v$. 

\begin{figure}[t]
\centering
\includegraphics[width=.975\columnwidth]{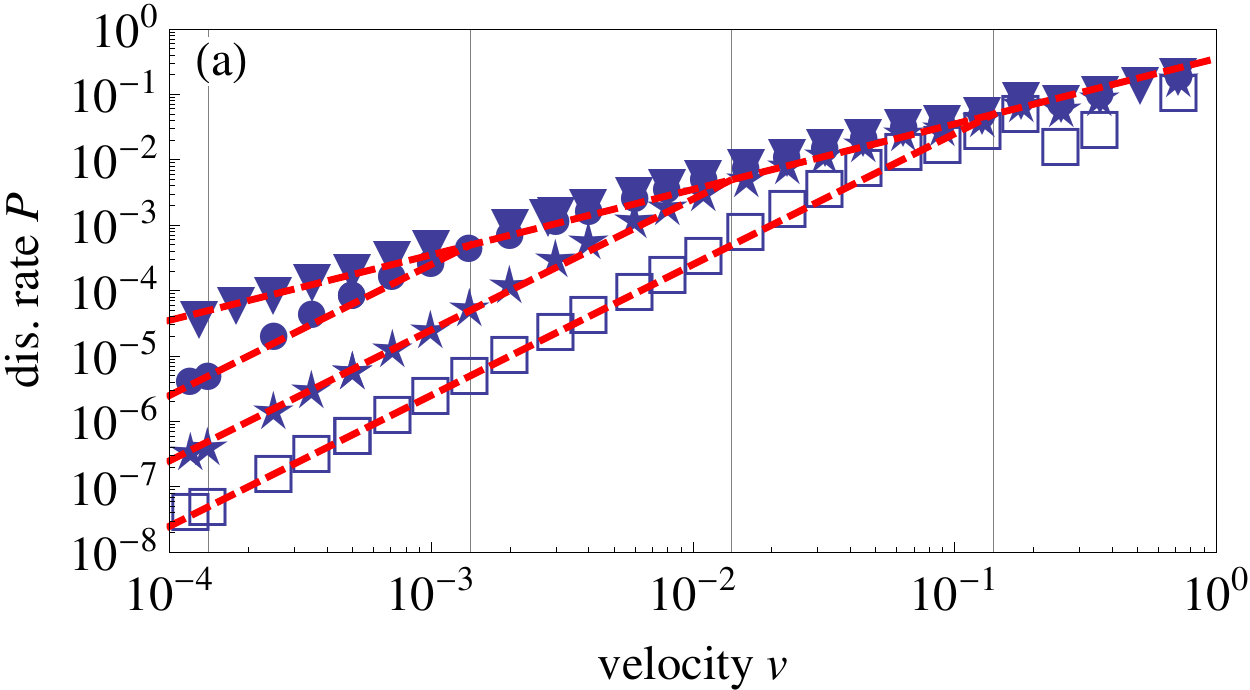}
\includegraphics[width=.975\columnwidth]{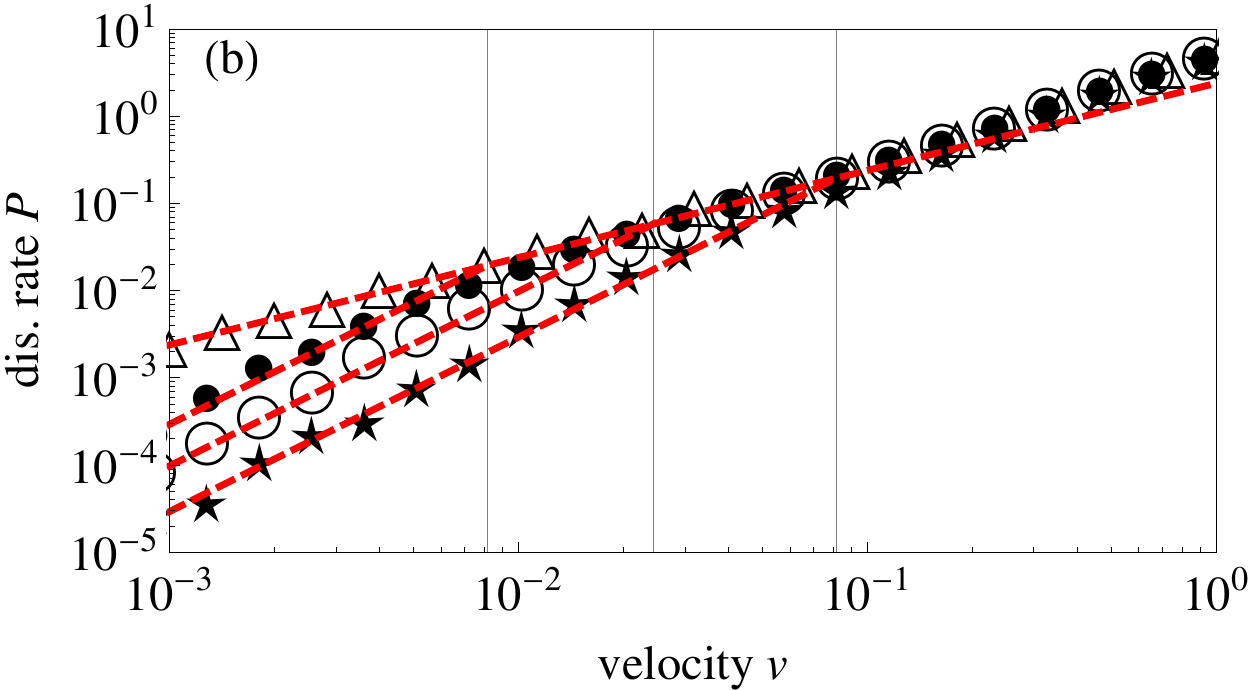}
\includegraphics[width=.975\columnwidth]{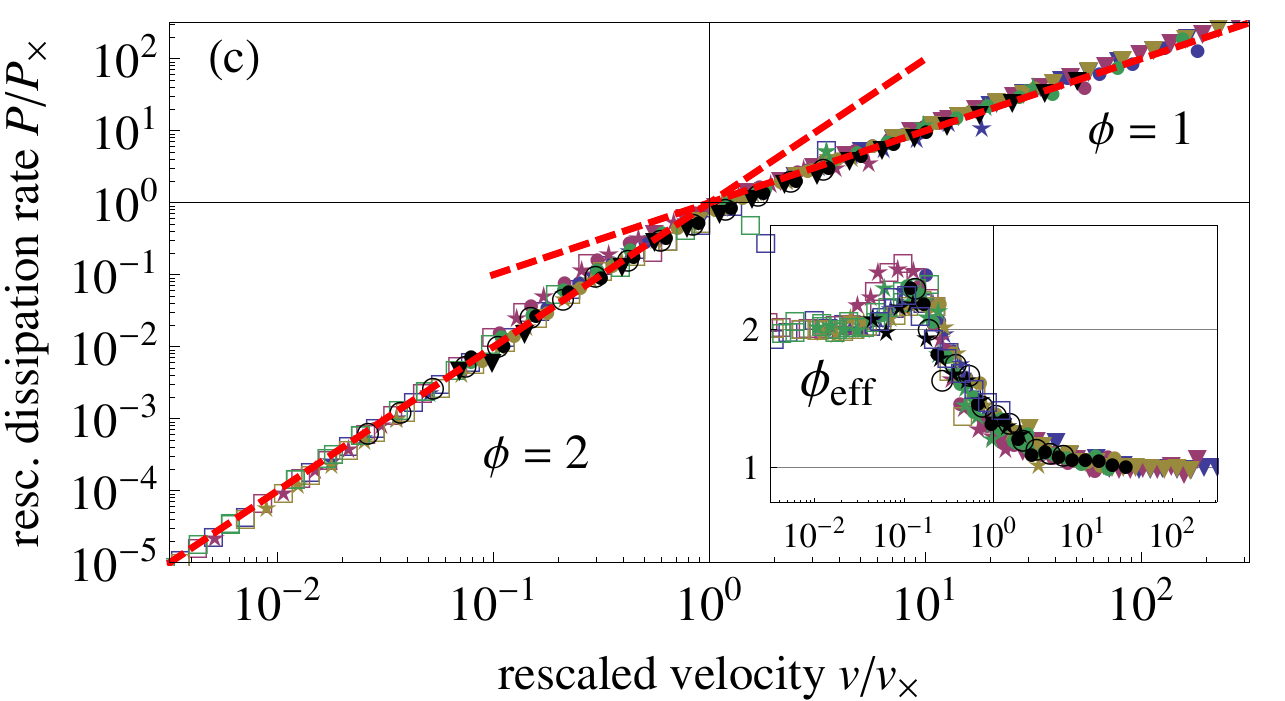}
\caption{(Color online) Dissipated power vs.\ velocity (in natural units) for the a) Heisenberg ($d_x {=} 0.5J, \alpha {=} 0.5,
h{=}J$, blue) and the b) Ising ($\beta {=} 1/J, h{=}10J$, black) model.
The simulated $\delta r$ are $\delta r{=}\infty$ ($\vartriangle$), $10^{-4}$
($\blacktriangledown$), $10^{-3}$ (\textbullet), $3 \times 10^{-3}
(\circ)$, $10^{-2}$ (\textborn) and $10^{-1}$ ($\square$). The grid
lines mark the corresponding $v_\times$, the dashed lines display the calculated
$P_\mathrm{C}(v)$ from Eq.~(\ref{eq:PpumpCo}) and fitted
$P_\mathrm{S}(v)$.
For c) we calculated explicitly the cross-over quantities $P_\times$ and
$v_\times$ for both models from
Eqs.~(\ref{eq:PcEqualsPh},\ref{eq:vcHe}), and plot the data again rescaled. Additionally we varied: $\alpha {=} 0.3$ (purple),
$h{=}2J$ (green) and $d_x{=}0.25J$ (yellow)  for the same $\delta r$ set. 
In the inset we plot an effective exponent $\phi_\mathrm{eff}{=} \partial \log
P/\partial \log v$, and get a universal cross-over from $2$ to $1$.
\label{fig:crossover}
}
\end{figure}

We first consider a chain of classical, normalized Heisenberg spins ($|\mathbf S_r| {=}
1$) of length $L$ with lattice spacing $a$, which interact with the
field defined above. The
corresponding  time-dependent Hamiltonian is
\begin{equation}
  \mathcal H (t)= -\sum_{r=1}^{L} J\; \mathbf S_r \cdot \mathbf
    S_{r+1} + d_x S_{r,x}^2 +  h_z(r-vt)  S_{r,z},
\label{eq:Hamiltonian}
\end{equation}
with the exchange constant $J$.
To get a well defined ground
state, we use an easy axis anisotropy ($d_x{>}0$) and anti-periodic boundary conditions $\mathbf S_{r+L}{=}{-}\mathbf
S_r$.  The spins perform Landau-Lifshitz-Gilbert
dynamics \cite{LandauLifshitz1935, Gilbert1955},
\begin{equation}
\frac {\mu_s(1+\alpha^2)}{\gamma}  \frac{\partial \mathbf S_r}{\partial t} = 
    \mathbf S_r {\times} \frac{\partial \mathcal H}{\partial \mathbf S_r} + \alpha
    \mathbf S_r \times 
    \left ( \mathbf S_r {\times} \frac{\partial \mathcal H}{\partial \mathbf S_r}
    \right),
\label{eq:LLG}
\end{equation}
consisting of a precessional motion with a frequency proportional to
$\gamma/\mu_s$, and a damping with the damping constant $\alpha$. For
simplicity, we neglect temperature here, and the dynamic parameters
yield a spin relaxation time $\tau_\mathrm{rel}$.  The friction force
$F$ can be either calculated from the dissipated power $P_\mathrm{diss}$ or the
pumping power $P_\mathrm{pump}$, which are equal in the stationary state due to
energy conservation and therefore we subsequently use $F=\left <
  P\right >/v$ after time averaging. The two cases can be described by
\begin{equation}
P(v)\propto  v^\phi,
\end{equation}
with the dissipation exponent $\phi{=}1$ ($\phi{=}2$) for the Coulomb
(Stokes) case. $P$ can be extracted from the
energy terms by
\begin{equation}
P(t) = \frac{\partial \mathcal H}{\partial t} =
\sum_{r=1}^L{\frac{\partial h_z(r-vt)}{\partial t}   S_{r,z}} 
\label{eq:Pt}
\end{equation}
which represents the power pumped into the system by the motion.  In
our simulations (c.f.\ Fig.~\ref{fig:crossover}a) we found $\phi{=}2$ for
large $\tau_\mathrm{switch}$, which corresponds to the results in
\cite{Fusco08, Magiera09, Magiera11, NIC-Proceeding}. For sufficiently small
$\tau_\mathrm{switch}$ we get $\phi{=}1$, which was
known from simulations in the Ising model, and was now reproduced in
the Heisenberg model.

\begin{figure}[t]
\includegraphics[width=.95\columnwidth]{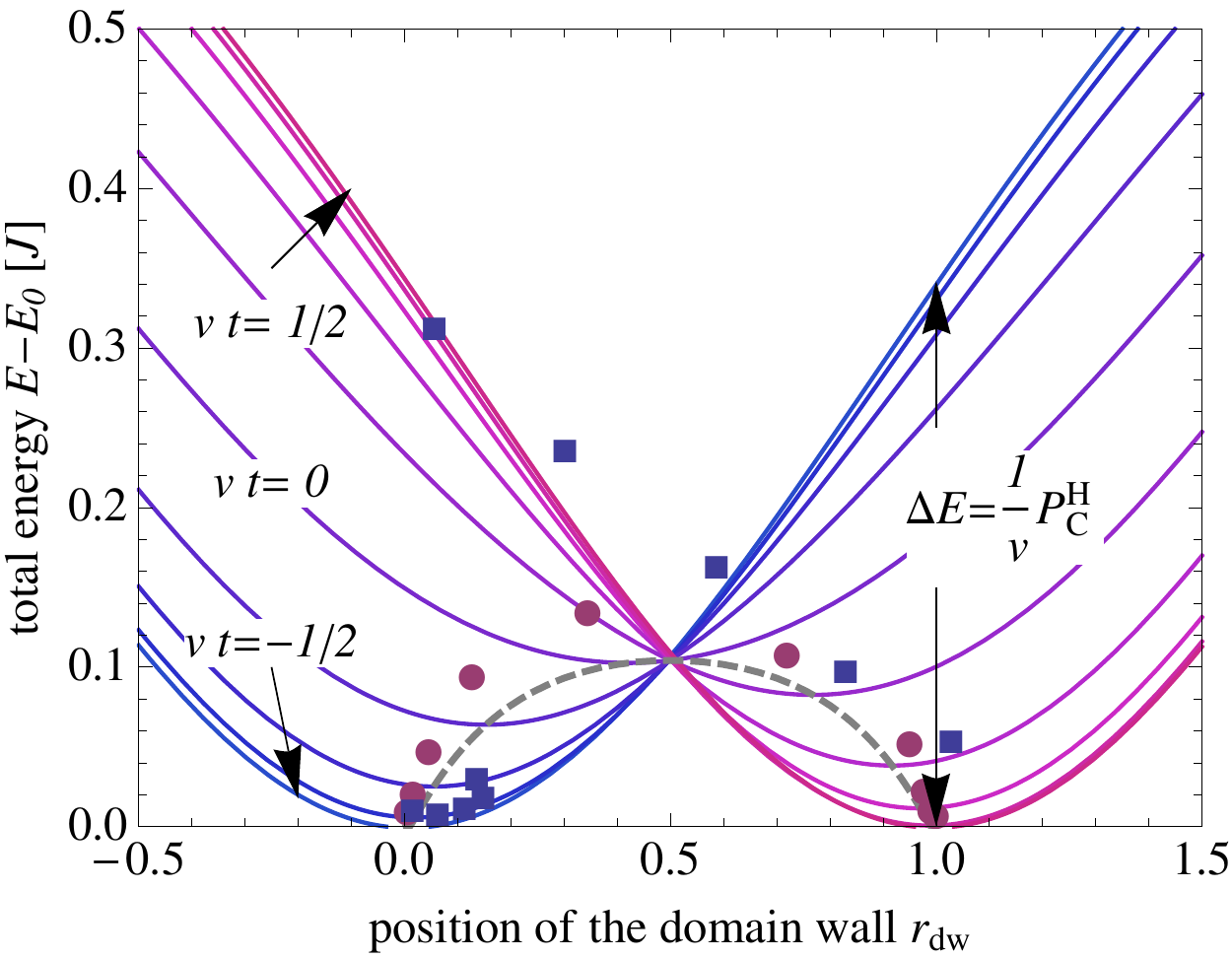}
\caption{\label{fig:potential}The system can be parameterized by the
  center of the DW $r_\mathrm{dw}$, thus for different times
  the total energy of the system can be calculated.  This is done for
  field parameter $\delta r {=}0.1$ for 10 equidistant times (blue to
  purple curves). If the system evolved quasi static, it would follow the
  configuration of minimal energy, marked by the 
  curve. Simulation results show both cases, the points indicate energy
  vs.\ $r_\mathrm{dw}$ for 10 time steps:
 (i)
  $\blacksquare$ The energy $P_\mathrm{C}^\mathrm{H}/v $ is
  periodically pumped into the system, which relaxes independent from
  $\tau_\mathrm{switch}$ afterwards. As the switching occurs at
  $vt=0$, we first see the relaxation from the preceding excitation in
  the left minimum, and at $vt=1/2$ the relaxation in the right
  minimum is not finished.
 (ii) $\bullet$ The
  system follows with a lag, but stays near equilibrium, slightly above the minimal energy configuration.
}
\end{figure}

In the following we calculate the velocity $v_\times$ at
which a cross-over from one regime to the other occurs.
For case (i) only two spins contribute to the sum in Eq.~(\ref{eq:Pt})
at the discrete times $vt \in \mathbb{Z}$ (at all other times and positions the
field remains constant), and we can calculate the averaged
pumping power by discretizing $\partial_t h_z$,
\begin{equation}
  \label{eq:PpumpCo}
  P_{\mathrm{C}} = -hv \left  (
    \left < S_{1',z} \right >- \left < S_{0',z} \right > \right ).
\end{equation}
For the time
$\tau_\mathrm{ca}$, corresponding to the time at which the amplitude
of the field stays
nearly constant, no pumping (respectively excitation) occurs.
We consider $\tau_\mathrm{ca}{>}\tau_\mathrm{rel}$, i.e.\ the system
always relaxes to equilibrium after a pumping event. Since the
equilibrium configuration does not depend on the dynamics,
Eq.~\ref{eq:PpumpCo} tells that here $\phi {=} 1$. 
The equilibrium configuration for our choice of boundary conditions is
a domain wall (DW) state, where the out-of-axis component is determined by
the field and thus points in $z$-direction. As the field interacts
mainly with only one spin, the shape of the DW is not
influenced by $h$
and we may use the continuum limit profile ($a {\rightarrow} 0$)
\footnote{Quantities with the subscript \textit{H} are dedicated
  to the Heisenberg model, those with the subscript \textit{I} to
  the Ising model.}
\begin{equation}
\mathbf m^\mathrm{H}(r') = \left (\tanh{\left (r'/\ell \right )},  0, \mathrm{sech}{\left  (r'/\ell \right )} \right),
\label{eq:Domainwall}
\end{equation}
with the DW width $\ell {=} \sqrt{J/(2 d_x)}$, which can be calculated from minimizing the free energy \cite{Bulaevskii63}. By inserting
$\langle \mathbf S_{0'} \rangle {=}\mathbf m^\mathrm{H}(0)$ and
$\langle \mathbf S_{1'} \rangle {=}\mathbf m^\mathrm{H}(1)$ into
Eq.~(\ref{eq:PpumpCo}) we now can calculate the power which is pumped
into the system during each switching event. This quantity can be
visualized in a potential plot. We again assume that $h$ does
not influence the shape of the DW but its center $r_\mathrm{dw}$. As in limiting case (ii) the
system is always near equilibrium, and in limiting case (i) it always
reaches the ground state before being excited out of equilibrium, this assumption is justified and we
can describe the whole configuration with $r_\mathrm{dw}$.  We look at
one cycle at which the field's peak moves from $0$ to $1$,
corresponding to the times $-1/2 {\le} vt {\le} 1/2$. For given $t$ we
can calculate the system's total energy as a function of
$r_\mathrm{dw}$ (see the potential lines in Fig.~\ref{fig:potential}).  If the
system evolved quasi statically, it would always be in the current
potential minimum. In this picture $P_\mathrm{C}/v$ corresponds to
the energy difference between the energy at $r_\mathrm{dw}{=}0, \,\, v
t {=} {-}1/2$ (the equilibrium state) and $v t{=}1/2$ (the state which
is present when the peak of the field has moved to the next spin while
the DW is still at the same site). Results from simulations
(plotted as squares in Fig.~\ref{fig:potential}) confirm this: At $v t {=} 0$
the system is excited to the upper state in a short time, and relaxes
to the new ground state by adjusting $r_\mathrm{dw}$ slowly
afterwards, until it reaches the new ground state configuration with
$r_\mathrm{dw}{=}1$.
Simulations of the second limiting case (circles in Fig.~\ref{fig:potential})
confirm that the system is always near equilibrium, thus the DW
slightly lags behind the ground state.
From our simulations, we found the pumping power 
\begin{equation}
P_\mathrm{S}^\mathrm{H} = \frac{d_x \mu_s}{J \gamma \delta r}
\alpha v^2.
\label{eq:PStokesHe}
\end{equation}
The factor $d_x v/(\delta r J)$ originates in the synchronization with
the field, which changes at the time scale $\tau_\mathrm{switch}$. The
factor $\alpha v \mu_s/\gamma$ emerges from spin dynamics, yielding a
retardation of the DW as derived in Ref.~\cite{Magiera09}. Setting
\begin{equation}
  \label{eq:PcEqualsPh}
  P_\mathrm{C}^\mathrm{H}(v) \stackrel{!}{=} P_\mathrm{S}^\mathrm{H} (v) := P_\times ^\mathrm{H} (v_\times ^\mathrm{H})
\end{equation}
yields the cross-over velocity
\footnote{The  unit of time is $\mu_s/(\gamma J)$. $d_x$ and $h$ are energies and given in units of $J$, thus the velocity's unit is $a\gamma J/\mu_s$.}
\begin{equation}
v^\mathrm{H}_\times=\frac{h \delta r}{\alpha d_x} \frac{\gamma
  J}{\mu_s} \left  (1-\mathrm{sech}\left (1/\ell\right )
 \right),
\label{eq:vcHe}
\end{equation}
where the system performs a cross-over from the Stokes-friction state to the
Coulomb-friction state.
In Fig.~\ref{fig:crossover}c these cross-over quantities
have been calculated and the simulation results have been rescaled
appropriately. The simulation data fit excellent over several
magnitudes with the derived cross-over quantities, remaining deviations are discussed below.
We performed also simulations of the isotropic Ising model
with the same
field and periodic boundary conditions ($S_{r+N} {=} S_{r}$).
The Ising spins undergo spin flip dynamics with
Metropolis probability \cite{Metropolis53}: Randomly chosen spins are
flipped with the probability $p_f {=} \min{(1,\exp{(-\beta \Delta
    E)})}$, where $\beta$ is the inverse temperature and $\Delta E$
the energy difference between the flipped and the not flipped
state.

For (i) we again find a behavior $\phi{=}1$.
We assume that for (i) the spins relax after each switching event to the
ground state profile which can be calculated via transfer matrix
  methods \cite{Hucht09}:
\begin{equation}
  m^{\mathrm{I}}(r') =\tanh{(\beta
    h)}[\tanh{(\beta J)}]^{|r'|}.
  \label{eq:Ismagprof}
\end{equation}
With $\langle S_{0',z} \rangle {=}m^\mathrm{I}(0)$ and $\langle
S_{1',z} \rangle {=}m^\mathrm{I} (1)$ in Eq.~(\ref{eq:PpumpCo}),
we get $P_\mathrm{C}^{I}$. $\phi{=}2$ is observed for (ii), and we
fitted
\begin{equation}
P_\mathrm{S}^\mathrm{I} \propto v^2/\delta r.
\end{equation}
We calculated again the cross-over velocity $v_\times^\mathrm{I}$, which is additionally
plotted in Fig.~\ref{fig:crossover}b, and rescaled all data points for
the cross-over plot, Fig.~\ref{fig:crossover}c.

Comparing Fig.~\ref{fig:crossover}a and Fig.~\ref{fig:crossover}b we
come to the main result of our investigation, namely coincidence
concerning the cross-over between both models, despite the substantial
remaining differences like e.g.\ the dynamics of the models.  The
present deviations from the cross-over curve are discussed below. 
The slight increase of $P$,
observed in the regime $v {>} 0.1$ for all $\delta r$ in the Ising model is due to the fact
that the system has not enough time to relax back to equilibrium
before the next shift takes place and $ m^\mathrm{I}(1)$ becomes
significantly smaller than its equilibrium value. 
As the Heisenberg model contains spin wave excitations, we observed the generation of spin waves above a
threshold velocity \cite{Magiera11}. In the
cross-over plot these spin waves cause a kink above
$v/v_\times{=}0.1$ for $\alpha{=}0.3$ (and a higher peak in the effective exponent plot). 
For very high velocities we observe a lowering of the power, which is due to a segregation
of the peak of the field and the DW, leading to a reduced
$m^\mathrm{H}_z(0){<}1$.  This state with lowered dissipation has already
been observed and reported in Ref.~\cite{NIC-Proceeding}.

In conclusion, we presented a new model, which for the case of magnetic
friction shows a transition from Stokes to Coulomb behavior, analogous
to the Prandtl-Tomlinson model for solid friction. Whereas there, the
elastic stiffness of the slider was the crucial parameter, it is the
switching time of the magnetic field in our case. The comparison of
both models sheds new light on the univeral origin of Coulomb
behavior, which is based on a separation of the relaxation time from
the much larger time scale on which the system gets excited.
Our findings are in accordance to field theoretical results
by Demery et al.\ \cite{Demery10, Demery10_2}, who also found
Stokes-like friction as their system model does not contain discrete
sites and thus $\tau_\mathrm{ca}=0$, i.e.\ the field is continiously
driving the system. However, their simulation results are not correct as they simulated an Ising
model with a discontinuous motion of a field, which is known to show
Coulomb friction. This discrepancy stems from an incorrect
definition of the friction force (Eq.~(50) in Ref.\cite{Demery10_2},
 a correct definition has been presented in \cite{Kadau08}).
\begin{acknowledgments}
  This work was supported by the German Research Foundation (DFG) via
  SFB 616 and the German Academic Exchange Service (DAAD) through
  the PROBRAL programme.
\end{acknowledgments}

\end{document}